\documentclass[doublecol]{epl2}
\usepackage{graphicx}
\usepackage{bm}
\usepackage{amssymb}
\usepackage{amsmath}
\usepackage[colorlinks=true,linkcolor=blue,citecolor=blue,urlcolor=blue]{hyperref}
\newcommand{\be}{\begin{equation}}
\newcommand{\ee}{\end{equation}}
\newcommand{\beqn}{\begin{eqnarray}}
\newcommand{\eeqn}{\end{eqnarray}}

\title{Quantum relaxation and finite size effects in the XY chain in a transverse field after global quenches}
\shorttitle{Quantum relaxation in the XY spin chain\dots} 

\author{B. Bla\ss\inst{1}\thanks{E-mail: \email{bebla@lusi.uni-sb.de}}, H. Rieger\inst{1}\thanks{E-mail: \email{h.rieger@mx.uni-saarland.de}} \and F. Igl\'oi\inst{2,3}\thanks{E-mail: \email{igloi.ferenc@wigner.mta.hu}}}

\shortauthor{B. Bla\ss \etal}

\institute{
  \inst{1} Theoretische Physik, Universit\"at des Saarlandes, 66041
  Saarbr\"ucken, Germany\\ \inst{2} Wigner Research Centre, Institute
  for Solid State Physics and Optics, H-1525 Budapest, P.O.Box 49,
  Hungary\\ \inst{3} Institute of Theoretical Physics, Szeged
  University, H-6720 Szeged, Hungary }

\date{\today}

\abstract{We consider global quenches in the quantum XY chain in 
a transverse field and study the nonequilibrium relaxation of the
magnetization and the correlation function as well as the entanglement
entropy in finite systems. For quenches in the ordered phase the
exact results are well described by a semiclassical theory (SCT) in
terms of ballistically moving quasi-particle pairs. 
For finite systems quasi-periodic behaviour of the dynamical 
evolution of the local order parameter and the correlation functions is 
predicted correctly including the period length, an exponential relaxation, 
a quasi-stationary regime and an exponential recurrence in one period. In 
the thermodynamic limit the SCT is exact
for the entanglement entropy and its modified version following the
method of \textit{Calabrese, Essler and Fagotti: J. Stat. Mech. (2012) P07016} is
exact for the magnetization and the correlation function, too. The
stationary correlation function is shown to be described by a
generalized Gibbs ensemble.}


\begin{document}
\maketitle

\section{Introduction}

Recent progress of experimental work on ultracold atomic gases in
optical lattices
\cite{Greiner_02,Paredes_04,Kinoshita_04,Kinoshita_06,Lamacraf_06,
Sadler_06,Hofferberth_07,Trotzky_12,Cheneau_12,Gring_11}
has made studies of the unitary time evolution of quantum systems in
nonequilibrium situations possible, \emph{i.e.}, after so-called global quenches. This
is usually achieved by a sudden change of the parameters of the
quantum system within a time scale that is much shorter than the
characteristic time the system needs to relax into a stationary state.
The basic questions in this context are i) the dynamical
characteristics of the relaxation process and ii) the possible
existence of a stationary state after long times. In three-dimensional
systems fast relaxation into a thermal stationary state has been
observed. In contrast to this, in quasi-one-dimensional systems the
relaxation process has been found to be much slower and to lead to an
unusual nonthermal stationary state\cite{Kinoshita_06}. This result
has stimulated intensive theoretical
work\cite{Polkovnikov_11,Rigol_07,Calabrese_07,Cazalilla_06,Manmana_07,
Cramer_08,Barthel_08,Kollar_08,Sotiriadis_09,Roux_09,Sotiriadis_11,
Kollath_07,Banuls_11,Gogolin_11,Rigol_11,Caneva_11,Cazalilla_11,Rigol_12,
Santos_11,Grisins_11,Canovi_11}
to clarify the effect of integrability of the system on the relaxation
process as well as on the nature of the stationary state. It is
commonly expected that observables of nonintegrable systems
effectively thermalize, which means that their stationary state is
described by a thermal Gibbs ensemble. Numerical studies of different
non-integrable systems are in favour of this
expectation\cite{Rigol_07,Calabrese_07,Cazalilla_06,Manmana_07,
Cramer_08,Barthel_08,Kollar_08,Sotiriadis_09,Roux_09,Sotiriadis_11},
however some contradictory results indicate that the issue could be
more complicated\cite{Kollath_07,Banuls_11,Gogolin_11,Grisins_11}. On
the other hand in integrable systems, due to the existence of
integrals of motion, the stationary state is expected to be
represented by a generalized Gibbs ensemble (GGE)\cite{Rigol_07}, in
which each mode corresponding to a conserved quantity is characterized
by its own effective temperature. Results on integrable systems are
mainly collected on free-fermion models, such as on the transverse
Ising chain, for which several analytical and numerical results have
been recently
obtained\cite{Barouch_70,Igloi_00,Sengupta_04,Fagotti_08,Silva_08,Rossini_09,
Campos_Venuti_10,Igloi_11,Foini_11,Rieger_11,Calabrese_11,Schuricht_12,
Calabrese_12,Calabrese_12b}. Qualitative features of the relaxation
process can be explained with a quasi-particle (QP)
picture\cite{Calabrese_05,Calabrese_07}: The quench changes the total
energy of the system by an extensive amount, which creates QPs
homogeneously in space, moving ballistically with a constant
velocity. Due to the conservation of momentum, the QPs are created in
pairs with opposite velocity and these QP pairs are quantum
entangled. This QP picture has been used to explain the time evolution
of the entanglement entropy\cite{Fagotti_08,Eisler_09} and has been
made quantitative with a semiclassical (SC) theory to predict also
the relaxation of the local magnetization and correlation function
\cite{Igloi_11,Divakaran_11,Rieger_11}. 
The basic idea of this
SC theory is, that the QPs can be identified with kinks or
domain walls in the spin chain that are created through the quench and
then move uniformly via the action of the $\sigma^z$ operator in the
transverse field term on states in the $\sigma^x$ representation. A
priori there is no reason to expect a similar mechanism to be at work
when the Ising symmetry is missing, as for instance in the XY model, in
particular when the transverse field is absent. In this letter we
will show that a quantitative description of the relaxation process
with uniformly moving kink pairs is also applicable for finite and
infinite XY chains. The main reason is that the $\sigma^y\sigma^y$
operator in the XY model has a similar effect on states in the
$\sigma^x$ representation as the $\sigma^z$ operator in the TIC,
namely creation and translation of kinks, in this case not by 1 but by
2 lattice spacings.

Therefore we study in this letter the XY chain in a transverse field
(denoted simply as XY chain). This model is integrable and has been
studied in detail both in equilibrium\cite{Lieb_61,Barouch_70} and out
of equilibrium\cite{Barouch_70,Fagotti_08} after a global quench. 
Here
we consider large finite chains and calculate the time dependence of
the entanglement entropy, the local magnetization and the equal-time
correlation function by free-fermion
techniques\cite{Lieb_61,Barouch_70}. These results are then compared
with the prediction of our SC theory. We show that in the
thermodynamic limit the SC theory provides exact results for the
entanglement entropy. For the magnetization and for the equal-time
correlation function the SC theory can be modified along the lines of
ref.\cite{Rieger_11}, so that it will become asymptotically
exact\cite{Calabrese_11,Calabrese_12}, too.



\section{Model and the free-fermion representation}

The XY chain is defined by the Hamiltonian
\be
{\cal H}=-\tfrac{1}{2}\sum_l \left[\tfrac{1+\gamma}{2} 
\sigma^x_l \sigma^x_{l+1}+\tfrac{1-\gamma}{2} \sigma^y_l \sigma^y_{l+1}\right]
-\tfrac{h}{2} \sum_l \sigma^z_l
\label{hamilton}
\ee
in terms of the Pauli spin operators $\sigma_l^{x,y,z}$ at site
$l$.
Generally, we consider large open
chains of length $L$.
The parameters $0 \le \gamma \le 1$ and $h \ge0$
denote the strength of the anisotropy and the transverse field,
respectively. The special case $\gamma=1$ represents the transverse
Ising model, and for $h=0$, $\gamma=0$ the Hamiltonian reduces
to the XX chain (see the equilibrium phase diagram in
fig.\,\ref{Fig_phase_diagram}).

We consider global quenches (at zero temperature), which suddenly
change the parameters of the Hamiltonian from $\gamma_0$, $h_0$ for
$t<0$ to $\gamma$, $h$ for $t>0$. For $t<0$ the system is assumed
to be in equilibrium, \emph{i.e.}, in the ground state of the Hamiltonian
${\cal H}$ with parameters $\gamma_0$ and $h_0$, which is denoted by
$\left|\Phi_0\right\rangle$. After the quench, for $t>0$, the state
evolves coherently according to the new Hamiltonian as
$\left|\Phi_0(t)\right\rangle=\exp(-\imath{\cal
H}t)\left|\Phi_0\right\rangle$.  Correspondingly the time evolution of
an operator in the Heisenberg picture is
$\sigma_l\left(t\right)=\exp\left(\imath{\cal H}t\right) \sigma_l
\exp\left(-\imath{\cal H}t\right)$.

We calculate the equal-time correlation function
$C^{xx}_t\left(l_1,l_2\right)=\left\langle \Phi_0\left|\sigma^{x}_{l_1}(t) 
\sigma^{x}_{l_2}(t) \right|\Phi_0\right\rangle$
, which for large separations is given by
$C^{xx}_t\left(l_1,l_2\right)=m_{l_1}\left(t\right)m_{l_2}\left(t\right)$,
where $m_l\left(t\right)$ is the local magnetization.  In the initial
state in the thermodynamic limit
one has $m_l\left(0\right)>0$ ($m_l\left(0\right)={\cal
O}\left(1/L\right)$) for $h_0<1$ ($h_0>1$) and at $h_0=1$ there is a
quantum critical line, which belongs to the (transverse) Ising
universality class\cite{Yang_52} for $\gamma>0$,
\emph{i.e.}, $m_l\left(0\right) \sim L^{-1/8}$.

Using standard techniques\cite{Lieb_61}, the Hamiltonian in
eq.\,(\ref{hamilton}) is expressed in terms of fermion creation and
annihilation operators $\eta^{\dag}_p$ and $\eta_p$ as
\be
{\cal H}=\sum_p \varepsilon\left(p\right)\left(\eta^{\dag}_p \eta_p-\tfrac{1}{2}\right)\;,
\ee
where the sum runs over $L$ quasi-momenta and the $p$ values are determined by
the boundary condition: $0<p<\pi$ ($-\pi<p<\pi$) for free (periodic) chains.  
The energy of the modes is given by
\be
\varepsilon\left(p\right)=\sqrt{\gamma^2 \sin^2 p+\left(\cos p-h\right)^2}
\label{eps}
\ee
and the Bogoliubov angle $\varTheta_p$ diagonalizing the Hamiltonian
is given by $\tan\varTheta_p=\gamma\sin p/\left(\cos
p-h\right)$. The correlation function is written as a Pfaffian, which is then evaluated through the
determinant of an antisymmetric matrix. The local magnetization is
calculated in the form of the off-diagonal matrix
element\cite{Yang_52} $m_l\left(t\right)=\left\langle
\Phi_0\left|\sigma_l^x\right|\Phi_1\right\rangle$, where
$\left|\Phi_1\right\rangle$ is the first excited state of the initial
Hamiltonian. 

The entanglement entropy between a block of $l$ contiguous spins and
the rest of the system is defined as ${\cal S}_l={\rm Tr}_{i \leq
l}[\rho_{l} \ln \rho_{l}]$, where $\rho_{l}= {\rm Tr}_{i >
l}|\Phi_0\rangle\langle\Phi_0|$ is the reduced density matrix, which
evolves in time as $\rho_{l}(t)=\exp(\imath{\cal H}t)
\rho_{l}\exp(-\imath{\cal H}t)$. 
For free-fermion models see the calculation in
ref.\cite{Igloi_09}.

\section{The semiclassical theory}

As mentioned in the introduction, QPs are created after the quench at
$t=0$. For the XY chain these QPs are the free-fermions described
above. The wave packets formed by the free-fermions move ballistically
with a constant velocity $\pm v_p$, which is obtained in the SC theory
as
\be
v_p=\dfrac{\partial\varepsilon\left(p\right) }{\partial p}=\dfrac{\sin p\left[h-\left(1-\gamma^2\right)\cos p\right]}{\varepsilon\left(p\right)}\;.
\label{v_p}
\ee
The position of the QP pairs at times $t>0$ can be easily calculated from their creation position and their
velocity; in finite open chains the QPs are reflected at the boundaries.

We also need the creation probability
$f_p=f_p\left(h_0,\gamma_0;h,\gamma\right)$ of the QP pair.  Here we
make use of the fact that in a homogeneous system the QPs are created
uniformly in space and that $f_p$ corresponds to the occupation
probability of mode $p$ in the initial state
$\left|\Phi_0\right\rangle$, thus $f_p=\left\langle \Phi_0\right|
\eta^{\dag}_p \eta_p \left|\Phi_0\right\rangle$. 
For the XY model it is expressed through the difference
$\Delta_p=\varTheta_p-\varTheta^0_p$ of the Bogoliubov angles as
$f_p=\tfrac{1}{2}\left(1-\cos \Delta_p\right)$ with
\be
\cos \Delta_p=\dfrac{\left(\cos p -h_0\right)\left(\cos p -h\right)+\gamma\gamma_0\sin^2 p}{\varepsilon\left(p\right)\varepsilon_0\left(p\right)}\;,
\label{Delta}
\ee
where the index $0$ refers to quantities before the quench.
We note that in open chains $f_p$ has a small position dependence near
the boundaries, which to leading order can be neglected for long chains.

Each pair of entangled QPs (and only those) with one
partner moving within the block and simultaneously the other in the
rest of the system, contributes an amount given by the
binary entropy $s_p=-\left(1-f_p\right)\ln\left(1-f_p\right)-f_p\ln
f_p$ to the entanglement entropy. Summing up the contributions of all
QP pairs, one obtains the value of the entanglement entropy at the
given time.

In the $\sigma^x$ representation the QPs represent
kink-like excitations. As described first for thermal excitations in
ref.\cite{Sachdev_97} and generalized afterwards for quantum quenches
in ref.\cite{Rieger_11}, a QP passing site $l$ changes the sign of the
local magnetization operator $\sigma^x_l$. If in a finite system the
same QP visits $l$ several times, $\sigma^x_l$ changes sign only if the
number of visits is odd. Summing up the contributions of all QPs which
have passed $l$ before $t$, one obtains the local magnetization
\cite{Rieger_11}
\be
m_l\left(t\right)=m_l(0)\cdot
\exp\left(-\frac{2}{\pi}
\int_0^\pi dp\,f_p(h_0,h)\,q_p(t,l)
\right)\;,
\ee
where for $l\le L/2$ 
and $t\le T_p=L/v_p$
\be
q_p(t,l)=
\left\{
\begin{array}{ccl}
v_p t & \;{\rm for}\; & t\le l/v_p\\
l     & \;{\rm for}\; & l/v_p<t\le (L-l)/v_p\\
1-v_p t & \;{\rm for}\; & (L-l)/v_p<t\le T_p
\end{array}
\right.
\ee
For $t>T_p=L/v_p$ $q_p$ is periodic: $q_p(t+nT_p,l)=q_p(t,l)$.
Similarly the correlation function can be
calculated \cite{Rieger_11}. 
For quenches deep in the ferromagnetic phase an
excellent agreement between the SC theory and the exact results is
obtained.  For quenches close to the critical point the agreement is
less good since here the kinks are not
sharply localized and the domain walls have a finite extent of the
order of the equilibrium correlation length. In the thermodynamic
limit this effect can be taken into account by using an 
effective occupation probability \cite{Rieger_11}
\be
f_p \to \tilde{f}_p=-\tfrac{1}{2}\ln\left|\cos \Delta_p\right|\;.
\label{f_tilde}
\ee
Since $\cos\Delta_p=1-2f_p$, one has 
$\tilde{f}_p=f_p+{\cal O}\left(f_p^2\right)$, implying $\tilde{f}_p=f_p$ 
for small QP density to first order.  For large QP density, \emph{i.e.}, for
large quenches, the replacement (\ref{f_tilde}) represents a
phenomenological improvement of the SC theory
and follows from an asymptotically exact evaluation of the
correlation function by Calabrese, Essler and Fagotti for the
transverse Ising chain\cite{Calabrese_11,Calabrese_12}. As a matter of
fact this evaluation works analogously for other free-fermion models,
such as for the XY chain. In the following we use this modified SC
theory also for finite chains in order to calculate the time evolution
of the magnetization and the correlation function .

\section{Results}

\begin{figure}
\begin{center}
\includegraphics[width=0.475\textwidth]{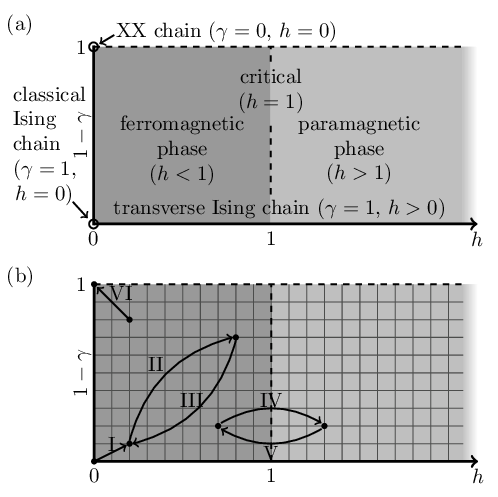}
\end{center}
\caption
{
\label{Fig_phase_diagram}
(a) Equilibrium phase diagram of the XY model at $T=0$, restricted 
to the anisotropy range $\gamma\in[0,1]$.
For $h<1,\,\gamma>0$ the system is ferromagnetic (FM) indicated by a
non-vanishing magnetization $m=\langle\sigma^x\rangle>0$ in the limit
$L\to\infty$, for $h>1$ the system is paramagnetic ($m=0$, PM). The
transition at $h=1$ between the FM and PM phase is of 2nd order and in
the universality class of the transverse Ising model (TIM)
$\left(\gamma=0\right)$. The point $\left(\gamma=1,\,h=0\right)$
represents the classical Ising model, the point
$\left(\gamma=0,\,h=0\right)$ is the XX chain, and the critical point
at $\gamma=0,\,h=1$ is in a universality class different from the
TIM. (b) Sketch of the different quenches $(h_0,\gamma_0\to
h,\gamma)$ considered in this Letter.  }
\end{figure}

We have performed quenches with six different pairs of parameters
$\left(h_0,\gamma_0\right) \to \left(h,\gamma\right)$ as indicated in
the phase diagram in fig.\,\ref{Fig_phase_diagram}.  Besides the
relatively small quench in the ordered phase (I) we have performed
larger quenches in the ordered phase (II and III) as well as quenches
between the ordered and the disordered phase (IV and V) and a quench
to the XX model (VI).

\subsection{Entanglement entropy}

\begin{figure}
\begin{center}
\includegraphics[width=0.43\textwidth]{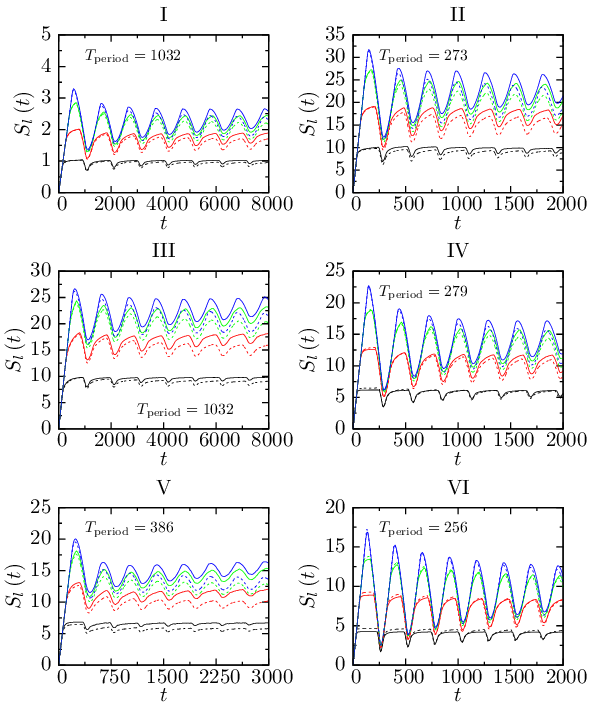}
\end{center}
\caption
{
\label{Fig_entropy} (Colour online) Time evolution of the entanglement entropy of the block of the first $l$ 
sites of the chain with free boundary conditions after the six quench protocols in fig.\,\ref{Fig_phase_diagram}.
Free-fermion (SC theory) results are indicated by full (broken) lines
(${\color{black}-}\,l=32$, ${\color{red}-}\,l=64$, ${\color{green}-}\,l=96$, ${\color{blue}-}\,l=128$).
}
\end{figure}

The dynamical entanglement entropy calculated by the free-fermionic method for a finite chain of
length $L=256$ and for various sizes of the block, $l$, are shown in fig.\,\ref{Fig_entropy}
together with the predictions of the SC theory. We observe an
excellent agreement for short times before the first
maximum, where the first QPs that have undergone a reflection at a 
boundary reach the middle of the chain. Small deviations are caused by
these reflections and accumulate in the subsequent periods.
In the SC calculation 
one should analyse the trajectory of the QP pairs
which for free boundary conditions
is summarized as follows. If the QP pair is
created at site $j$, then for $t<T_p=L/v_p$ the pair contributes to
the entropy within the interval $t_{p,1}<t<t_{p,2}$ with
$t_{p,1}=\left|l-j\right|/v_p$ and $t_{p,2}={\rm
min}\left[\left(l+j\right)/v_p,\left(2L-\left(l+j\right)\right)/v_p\right]$. 
For $T_p<t<2T_p$ this effective interval is $t_{p,3} <t<t_{p,4}$ with
$t_{p,3}=2T_p-t_{p,2}$ and $t_{p,4}=2T_p-t_{p,1}$. For $t>2T_p$ the
process is repeated with a period of $2T_p$.

As seen in fig.\,\ref{Fig_entropy}, the time evolution of the entropy
starts linearly, reaching a $l$-dependent maximum. Afterwards there is
a linear decrease and the process is repeated quasi-periodically as for
the transverse Ising chain\cite{Igloi_12} with a period $T_{\rm
period}=L/v_{\rm max}$, where $v_{\rm max}$ is the maximum group
velocity ${\rm max}_p\{v_p\}$. In the limits $L \to \infty$ and $l \gg 1$ 
the SC theory predicts
\be
S_l\left(t\right)=
\begin{cases}
t \frac{1}{2 \pi} \int_0^{\pi}dp \, v_p s_p,\quad & t<l/v_{max}\\
l \frac{1}{2 \pi} \int_0^{\pi}dp \, s_p,\quad & t \gg l/v_{max}\;,
\end{cases}
\ee
which corresponds to the exact results\cite{Fagotti_08}.


\subsection{Local magnetization}

\begin{figure}
\begin{center}
\includegraphics[width=0.475\textwidth]{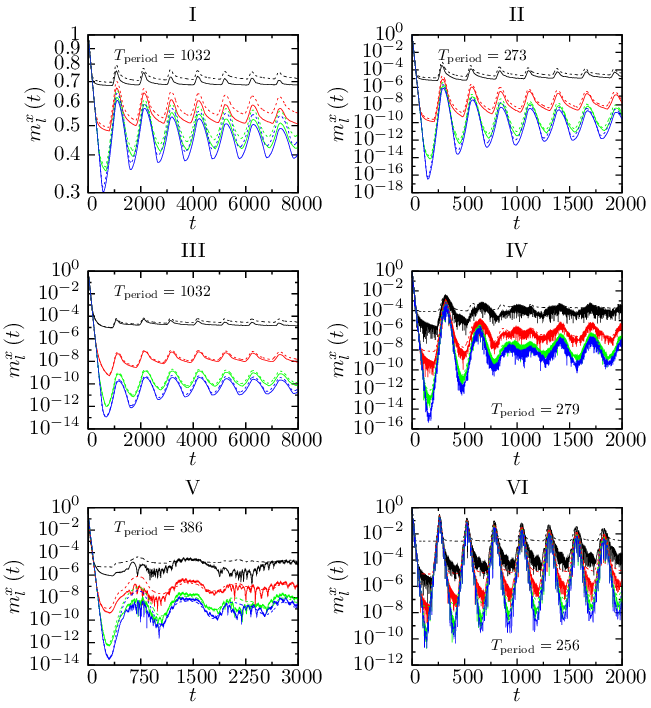}
\end{center}
\caption
{
\label{Fig_mag} (Colour online) Time evolution of the local 
magnetization $m_l\left(t\right)$ after the six quench protocols in
fig.\,\ref{Fig_phase_diagram}. Free-fermion (SC theory) results are
indicated by full (broken) lines (${\color{black}-}\,l=32$,
${\color{red}-}\,l=64$, ${\color{green}-}\,l=96$,
${\color{blue}-}\,l=128$). Note that the y-axis is
logarithmic such that straight curve segments represent either
exponential relaxation (negative slope) or exponential reconstruction
(positive slope).}
\end{figure}

The local magnetization
$m_l\left(t\right)$ calculated by the free-fermion method for finite open chains of length $L=256$
and at various positions
are shown in fig.\,\ref{Fig_mag}
together
with the predictions of the modified
SC theory for the six different quench protocols.
If the quench is performed between two
parameter points within the ordered phase, the modified SC theory
represents an excellent description of the relaxation process, in
particular for short times ($t<T_{\rm period}/2$). If the quench
involves the disordered phase as well or the XX point the agreement is
less good. The origin of the deviations here is the same as
for the entanglement entropy discussed above. However, the
qualitative features of the time evolution of the magnetization is
similar in all cases in fig.\,\ref{Fig_mag}: For finite systems 
the SC theory correctly predicts quasi-periodic behaviour including
the periodic time $T_{\rm period}$, an exponential relaxation, a
quasi-stationary regime and an exponential recurrence in one period.
These features have also been previously found for the the transverse 
Ising chain\cite{Igloi_11,Rieger_11}. 

In the limit $L \to \infty$ in the modified SC theory we have
\beqn
m_l\left(t\right)=m_l\left(0\right) \exp\left(-t\frac{2}{\pi}\int_0^{\pi} dp \,v_p \tilde{f}_p\theta\left(l-v_p t\right)\right.\nonumber\\
\left.-l\frac{2}{\pi}\int_0^{\pi} dp \, \tilde{f}_p \theta\left(v_p t-l\right) \right)
\label{m_l}
\eeqn
which defines the quench-dependent
relaxation time $\tau_{\rm mag}$ and the correlation length $\xi_{\rm
mag}$ as
\be
\tau_{\rm mag}^{-1}=\frac{2}{\pi}\int_0^{\pi} dp \,v_p \tilde{f}_p\;,
\quad \xi_{\rm mag}^{-1}=\frac{2}{\pi}\int_0^{\pi} dp \, \tilde{f}_p\;.
\label{tau_mag}
\ee

\subsection{Correlation function}

\begin{figure}
\begin{center}
\includegraphics[width=0.475\textwidth]{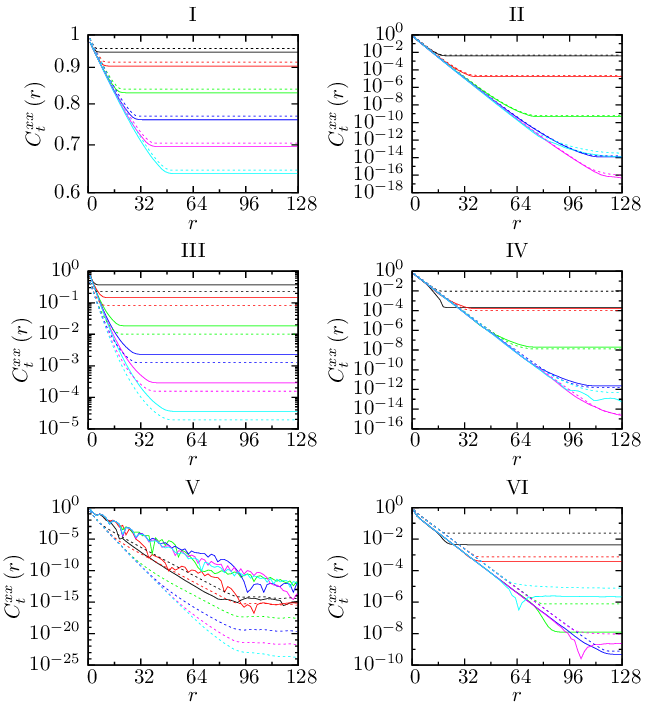}
\end{center}
\caption
{
\label{Fig_eqtc} (Colour online) Equal-time correlation function $C_t^{xx}\left(r\right)$ 
for fixed time $t$ after the six quench protocols in fig.\,\ref{Fig_phase_diagram} as
a function of the distance $r$. Free-fermion (SC theory) results are indicated by full 
(broken) lines (${\color{black}-}\,t=10$, ${\color{red}-}\,t=20$, ${\color{green}-}\,t=40$, 
${\color{blue}-}\,t=60$, ${\color{magenta}-}\,t=80$, ${\color{cyan}-}\,t=100$).
}
\end{figure}

The space-dependence of the equal-time correlation function,
$C^{xx}_t\left(r\right)=C^{xx}\left(\tfrac{L-r+1}{2},t;\tfrac{L+r+1}{2},t\right),\,r=1,3,\ldots,L-1$
between two sites symmetrically located between the boundaries of the
system calculated by the free-fermion method for $L=256$ and for various times 
is shown in fig.\,\ref{Fig_eqtc}
together with the predictions of the modified SC theory
for the same quenches
as indicated in fig.\,\ref{Fig_phase_diagram}.
Also in this case, an excellent agreement is seen
for short times before the stationary regime is observed. 
The agreement remains good for quenches between two parameter points in 
the FM phase. As for the magnetization, the agreement is less satisfactory 
if the quench involves also the disordered phase or the XX point. The origin of the
deviations here is the same as for the entanglement entropy. In the limit
$L\to\infty$ in the modified SC theory the correlation function is given by
\beqn
 C^{xx}_t\left(r\right)=C^{xx}_0\left(r\right) \exp\left(-t\frac{4}{\pi}\int_0^{\pi} dp \,
v_p \tilde{f}_p\theta\left(r-2v_p t\right)\right.\nonumber\\
\left.-l\frac{2}{\pi}\int_0^{\pi} dp \, \tilde{f}_p \theta\left(2v_p t-r\right) \right)\;,
\label{C_t}
\eeqn
%
which is analogous to the formula for the
local magnetization in eq.\,(\ref{m_l}). Note that the time $\tau_{\rm
corr}$ and the length scale $\xi_{\rm corr}$ of the correlation
function is related to that of the magnetization as $\tau_{\rm
corr}=\tau_{\rm mag}/2$ and $\xi_{\rm corr}=\xi_{\rm mag}$ as
for the transverse Ising chain\cite{Rieger_11}.

\section{Discussion}

We have studied the nonequilibrium relaxation dynamics of the quantum
XY chain after global quenches.  In particular, we have considered the
time evolution of the entanglement entropy as well as the relaxation
of the magnetization and the equal-time correlation function in finite
systems. The numerical results obtained with the free-fermion
techniques were compared with the predictions of the SC
theory. For the entanglement entropy an excellent agreement was found
for all types of quenches and the SC results were shown to be exact in
the $L \to \infty$ limit. For the magnetization and for the equal-time
correlation function the SC theory provides a very good approximation
if the quench is performed between two points deep within the ordered
phase. For two points in the ordered phase closer to the critical
line we showed that one can modify the SC theory by introducing an effective
particle occupation number, see in eq.\,(\ref{f_tilde}), which
provides exact results in the thermodynamic limit. In a finite system
new phenomena such as reconstruction of the magnetization and the
correlation function occur, which is explained in terms of
reflected QPs at the open boundaries of the chain or, for
periodic boundary conditions, in terms of the recurrence of the
QPs.

The XY chain after a sudden quench does not thermalize, since the QP
occupation probability cannot be described by a Gibbs distribution, or, more
concretely, there is no effective temperature $T_{\rm eff}$ depending
on the quench parameters, for which $f_p\sim\exp(-\varepsilon_p/T_{\rm
eff})$. Since all modes are conserved quantities, one rather expects 
that each mode has its own effective temperature $T_{\rm eff}(p)$,
which can be identified as follows:

When one compares the correlation length $\xi_{\rm corr}=\xi_{\rm
mag}$ in eq.\,(\ref{tau_mag}) with the thermal correlation length
$\xi_T$ at a finite temperature $T$\cite{Barouch_70}
\be
\xi_T^{-1}=-\dfrac{1}{\pi}\int_0^{\pi} dp \, \ln\left| \tanh \dfrac{\varepsilon\left(p\right)}{2T}\right|\;,
\label{xxx}
\ee
one observes in both cases an average over all modes. One can formally
obtain $\xi_{\rm corr}$ and $\xi_{\rm mag}$ from (\ref{xxx}) by
attributing a $p$-dependent effective temperature $T_{\rm eff}(p)$ to
the contribution of the $p$-mode, \emph{i.e.},
replacing the integrand in (\ref{xxx}) by
$\ln|\tanh(\varepsilon_p/2T_{\rm eff})|$. This then implies
\be
\tilde{f}_p=-\dfrac{1}{2}\ln\left|1-2f_p\right|=-\dfrac{1}{2}\ln\left| \tanh \dfrac{\varepsilon\left(p\right)}{2T_{\rm eff}\left(p\right)}\right|\;,
\ee
which leads to the relation
\be
{\rm min}(f_p,1-f_p)=\dfrac{1}{\exp\left( \frac{\varepsilon\left(p\right)}{T_{\rm eff}\left(p\right)}\right)+1 }\;.
\ee
Here on the right hand side there is the Fermi
distribution function, thus the nonequilibrium occupation probability
of the free-fermion mode $f_p$ (or that of the corresponding hole: $1-f_p$) is equal to its equilibrium thermal
occupation probability at $T_{\rm eff}\left(p\right)$. We note that
the same relation holds for the transverse Ising chain\cite{Rieger_11}
and it is expected to be valid for free-fermion models in
general. This supports the conclusion that after a sudden
quench the XY chain reaches a stationary state in which
correlations are described by a generalized Gibbs ensemble.

\acknowledgments
This work has been supported by the Hungarian National Research Fund
under grant Nos. OTKA K75324 and K77629.  F.I. is grateful to the
Theoretical Physics Department of the Saarland University for
hospitality at different stages of this work.

\end{document}